\documentstyle[12pt,epsfig]{article} 
\newcommand{\nd}{\noindent} 
\newcommand{\beq}{\begin{equation}} 
\newcommand{\eeq}{\end{equation}} 
\newcommand{\barr}{\begin{eqnarray}} 
\newcommand{\earr}{\end{eqnarray}} 
\newcommand{\ba}{\begin{array}} 
\newcommand{\ea}{\end{array}} 

\newcommand{\bfp}{\mbox{\boldmath $p$}} 
\newcommand{\bfP}{\mbox{\boldmath $P$}} 
\newcommand{\bfk}{\mbox{\boldmath $k$}} 
\newcommand{\bfy}{\mbox{\boldmath $y$}} 
 
\newcommand{\bfz}{\mbox{\boldmath $z$}}

\newcommand{\pup}{p^\uparrow} 
\newcommand{\aup}{a^\uparrow}

\newcommand{\adown}{a^\downarrow}

\newcommand{\Lup}{\Lambda^\uparrow} 
\newcommand{\Ldown}{\Lambda^\downarrow} 
\newcommand{\Blup}{\bar\Lambda^\uparrow} 
\newcommand{\Bldown}{\bar\Lambda^\downarrow} 
\newcommand{\hup}{h^\uparrow} 
\newcommand{\hdown}{h^\downarrow} 
\newcommand{\NP}[1]{{\it Nucl.\ Phys.}\ {\bf #1}} 
\newcommand{\ZP}[1]{{\it Z.\ Phys.}\ {\bf #1}} 
\newcommand{\EPJ}[1]{{\it Eur. Phys. J.}\ {\bf #1}} 
\newcommand{\PL}[1]{{\it Phys.\ Lett.}\ {\bf #1}} 
\newcommand{\PR}[1]{{\it Phys.\ Rev.}\ {\bf #1}} 
\newcommand{\PRL}[1]{{\it Phys.\ Rev.\ Lett.}\ {\bf #1}} 
\newcommand{\IJMP}[1]{{\it Int.\ J.\ Mod.\ Phys.}\ {\bf #1}} 
\newcommand{\MPL}[1]{{\it Mod.\ Phys.\ Lett.}\ {\bf #1}} 
 
\newcommand{\simorder}{\raisebox{-4pt}{$\, \stackrel{\textstyle >}{\sim} \,$}} 
\newcommand{\simordertwo}{\raisebox{-4pt}{$\, \stackrel{\textstyle <}
{\sim}\,$}} 
\def\lsim{\mathrel{\rlap{\lower4pt\hbox{\hskip1pt$\sim$}}\raise1pt\hbox{$<$}}} 
\def\gsim{\mathrel{\rlap{\lower4pt\hbox{\hskip1pt$\sim$}}\raise1pt\hbox{$>$}}} 
\def\nostrocostruttino#1\over#2{\mathrel{\mathop{\kern 0pt \rlap 
{\hbox{$#1$}}} \hbox{\kern-.135em $#2$}}} 
 
\begin{document} 
\begin{flushright} 
DFTT 32/2000 \\ 
INFNCA-TH0012 \\ 
hep-ph/0008186 \\ 
\end{flushright} 
\vskip 1.5cm 
\begin{center} 
{\bf {\mbox{\boldmath $\Lambda$}} 
polarization from unpolarized quark fragmentation}\\ 
\vskip 0.8cm 
{\sf M.\ Anselmino$^1$, D.\ Boer$^2$, U.\ D'Alesio$^3$ and F.\ Murgia$^3$} 
\vskip 0.5cm 
{$^1$ Dipartimento di Fisica Teorica, Universit\`a di Torino and \\ 
      INFN, Sezione di Torino, Via P. Giuria 1, I-10125 Torino, Italy\\ 
\vskip 0.5cm 
$^2$ RIKEN-BNL Research Center\\ 
Brookhaven National Laboratory, Upton, NY 11973, USA\\  
\vskip 0.5cm 
$^3$  Istituto Nazionale di Fisica Nucleare, Sezione di Cagliari\\ 
      and Dipartimento di Fisica, Universit\`a di Cagliari\\ 
      C.P. 170, I-09042 Monserrato (CA), Italy} \\ 
\end{center} 
\vskip 1.5cm 
\noindent 
{\bf Abstract:} \\  
The longstanding problem of explaining the observed polarization of 
$\Lambda$ hyperons inclusively produced in the high energy collisions of  
{\it unpolarized} hadrons is tackled by considering spin and $\bfk_\perp$ 
dependent quark fragmentation functions. The data on $\Lambda$'s  
and $\bar\Lambda$'s produced in $p-N$ processes are used to determine  
simple phenomenological expressions for these new  
``{\it polarizing fragmentation functions}'', which describe the experiments 
remarkably well. 
\newpage 
\pagestyle{plain} 
\setcounter{page}{1} 
\nd 
{\bf 1. Introduction} 
\vskip 6pt 
It is well known since a long time that $\Lambda$ hyperons produced with   
$x_F \simorder 0.2$ and $p_T \simorder$ 1 GeV/$c$ in the collision of two  
unpolarized  
hadrons, $A B \to \Lup X$, are polarized perpendicularly to the production  
plane, as allowed by parity invariance; a huge amount of experimental  
information, for a wide energy range of the unpolarized beams, is available  
on such single spin asymmetries \cite{data}: 
\beq 
P_\Lambda =  
\frac{d\sigma^{A B \to \Lup X} - d\sigma^{A B \to \Ldown X}} 
     {d\sigma^{A B \to \Lup X} + d\sigma^{A B \to \Ldown X}} \>\cdot 
\label{pol} 
\eeq 
Similar effects have been observed for several other hyperons, but we  
shall consider here only $\Lambda$'s and $\bar\Lambda$'s. 
 
Despite the wealth of data and the many years they have been known,  
no convincing theoretical explanation or understanding of the  
phenomenon exist \cite{theo1, theo2}. The perturbative QCD dynamics forbids 
any sizeable single spin asymmetry at the partonic level \cite{gol}; 
the polarization of hyperons resulting from the strong interaction 
of unpolarized hadrons must then originate from nonperturbative features, 
presumably in the hadronization process. A number of models attempting some  
understanding of the data in this perspective \cite{theo1}-\cite{bor} only  
achieve partial explanations. 
 
In the last years other single spin asymmetries observed in $\pup p \to 
\pi X$ reactions \cite{e704} have attracted a lot of theoretical  
activity \cite{siv}-\cite{giap}; a phenomenological description of such 
asymmetries appears now possible with the introduction of new  
distribution \cite{Ralst-S-79,siv,noi1,dan1} and/or fragmentation  
\cite{col,noi3,Mulders-Tangerman-96}  
functions which are spin and $\bfk_\perp$ dependent; $\bfk_\perp$ denotes 
either the transverse momentum of a quark inside a nucleon or of a hadron 
with respect to the fragmenting quark. 
 
In particular the effect first discussed by Collins \cite{col} -- that is, 
the azimuthal angle dependence of the number of hadrons produced in the  
fragmentation of a transversely polarized quark -- has been recently observed  
\cite{her,smc}; were such results confirmed the role of these new 
fragmentation functions would be of great phenomenological importance.  
 
We consider here an effect similar to that suggested by Collins, namely 
a spin and $\bfk_\perp$ dependence in the fragmentation 
of an {\it unpolarized} quark into a {\it polarized} hadron: a function  
describing this mechanism was first introduced in Ref.\  
\cite{Mulders-Tangerman-96} and denoted by $D_{1T}^\perp$. This   
function is introduced in a frame defined by two light-like four-vectors $n_+$ 
and $n_-$, satisfying $n_+\cdot n_-$ = 1, and by the plane transverse to  
them. The four-momentum $P$ of the outgoing hadron -- a $\Lambda$ hyperon  
in the present investigation -- is in the $n_-$ direction (up to a mass term   
correction). The function $D_{1T}^\perp$ is then defined as (displayed in the 
$n_+ \cdot A =0$ gauge)  
\barr 
\lefteqn{\frac{\epsilon_T^{ij} \, k_{Ti}^{} \, 
S_{Tj}^{}}{M_h} \, D_{1T}^\perp(z, k_\perp) \equiv 
\sum_X \int \frac{dy^+d^2\bfy_T^{}}{4z\,(2\pi)^3} \ 
e^{ik\cdot y}} \nonumber\\ 
&& \times \, {\mbox{${\rm Tr}$}} 
\left. \langle 0 \vert \psi (y) \vert P,S_T; X 
\rangle \langle P, S_T; X \vert 
\overline \psi(0) \gamma^- \vert 0 \rangle \right|_{y^- = 0}, 
\label{D1Tpdef} 
\earr 
where the final state depends on the transverse part ($S_T$) of the spin 
vector $S$ of the produced $\Lambda$ only, {\it i.e.}\ one should 
interpret it as $\vert P,S_T; X \rangle \equiv (\vert P,S=S_T; X 
\rangle - \vert P,S=-S_T; X \rangle)/2$, such that the 
contribution from unpolarized fragmentation cancels out. 
Furthermore, $k_\perp = |\bfk_\perp|$ is the modulus of the transverse 
momentum of the hadron in a frame where the fragmenting 
quark has no transverse momentum. More details on this type of definition 
of fragmentation (or decay) functions can be found in Refs.\ 
\cite{Coll-S-82,col,Mulders-Tangerman-96}.  
 
In the notations of Ref.\ \cite{noi3} a similar function is defined as: 
\barr 
\Delta^N D_{\hup/a}(z, \bfk_{\perp}) &\equiv& 
\hat D_{\hup/a}(z, \bfk_{\perp}) - \hat D_{\hdown/a}(z, \bfk_{\perp})  
\label{deld1}\\ 
&=& \hat D_{\hup/a}(z, \bfk_{\perp})-\hat D_{\hup/a}(z, - \bfk_{\perp}) \>, 
\nonumber 
\earr 
and denotes the difference between the density numbers  
$\hat D_{\hup/a}(z, \bfk_{\perp})$ and  
$\hat D_{\hdown/a}(z,$ $\bfk_{\perp})$ 
of spin 1/2 hadrons $h$, with longitudinal momentum fraction $z$, transverse  
momentum $\bfk_{\perp}$ and transverse polarization $\uparrow$ or  
$\downarrow$, inside a jet originated by the fragmentation of an  
unpolarized parton $a$. From the above definition it is clear that the 
$\bfk_{\perp}$ integral of the function vanishes.   
 
The exact relation between $D_{1T}^\perp$ and $\Delta^N D_{\hup/a}$ is 
given by (notice that also $D_{1T}^\perp$ should have labels $h$ and $a$
which are often omitted): 
\beq 
\Delta^N D_{\hup/a}(z, \bfk_{\perp})   
= 2 \frac{k_\perp}{z M_h} \; \sin\phi \;  
D_{1T}^\perp(z, k_\perp) \>, \label{rel}  
\eeq 
where $\phi$ is the angle between $\bfk_\perp$ and the  
transverse polarization vector of the hadron, which shows that the function  
$\Delta^N D_{\hup/a}(z, \bfk_{\perp})$ vanishes in case the transverse  
momentum and transverse spin are parallel. 
 
In the sequel we shall refer to $\Delta^N D_{\hup/a}$ and $D_{1T}^\perp$ 
as ``{\it polarizing fragmentation functions}''. 
 
In analogy to Collins' suggestion for the fragmentation of a transversely  
polarized quark we write \cite{col,noi4}: 
\beq  
\hat D_{\hup/q}(z, \bfk_\perp) = \frac 12 \> \hat D_{h/q}(z, k_\perp) +  
\frac 12 \> \Delta^ND_{\hup/q}(z, k_\perp) \>  
\frac{\hat{\bfP}_h \cdot (\bfp_q \times \bfk_\perp)} 
{|\bfp_q \times \bfk_\perp|} \label{lamfn} 
\eeq 
for an unpolarized quark with momentum $\bfp_q$ which fragments into  
a spin 1/2 hadron $h$ with momentum $\bfp_h = z \bfp_q + \bfk_\perp$ 
and polarization vector along the $\uparrow \> = \hat{\bfP}_h$ direction; 
$\hat D_{h/q}(z, k_\perp) =  D_{\hup/q}(z, \bfk_\perp) +   
D_{\hdown/q}(z, \bfk_\perp)$ is the $k_\perp$ dependent unpolarized  
fragmentation function. Notice that $\hat{\bfP}_h \cdot (\bfp_q \times  
\bfk_\perp) = \bfp_q \cdot (\bfk_\perp \times \hat{\bfP}_h) \sim \sin \phi$. 
  
A QCD factorization theorem gives for the high energy and large $p_T$  
process $A B \to \Lup X$, at leading twist with collinear parton  
configurations: 
\beq 
d\sigma^{AB \to \Lup X} = \sum_{a,b,c,d} f_{a/A}(x_a) \otimes  f_{b/B}(x_b) 
\otimes d\hat\sigma^{ab \to cd} \otimes D_{\Lup/c}(z) \label{dsu} 
\eeq 
and 
\beq 
d\sigma^{AB \to \Ldown X} = \sum_{a,b,c,d} f_{a/A}(x_a) \otimes  f_{b/B}(x_b) 
\otimes d\hat\sigma^{ab \to cd} \otimes D_{\Ldown/c}(z) \>. \label{dsd} 
\eeq 
Here and in the sequel we shall fix the scattering plane as the $x$--$z$ 
plane, with hadron $A$ moving along $+\hat{\bfz}$ and the detected $\Lambda$  
produced in the first $x$--$z$ quadrant; the $\uparrow$, $\downarrow$  
directions are then respectively $+\hat{\bfy}$ and $-\hat{\bfy}$. 
 
In the absence of intrinsic $\bfk_\perp$ (or rather when integrated over)  
the fragmentation functions $D_{\Lup/c}(z) = \int d^2 \bfk_{\perp} \>  
\hat D_{\Lup/q}(z, \bfk_{\perp})$ (or $D_{\Ldown/c}(z)$) 
cannot depend on the hadron polarization, so that one has  
$d\sigma^\uparrow = d\sigma^\downarrow$, which implies $P_\Lambda = 0$. 
 
However, by taking into account intrinsic $\bfk_\perp$ in the hadronization  
process, and assuming that the factorization theorem holds also when  
$\bfk_\perp$'s are included \cite{col}, one has, using Eq.\ (\ref{lamfn})  
instead of $D_{\Lambda/c}(z)$ in Eqs.\ (\ref{pol}), (\ref{dsu}) and (\ref{dsd}): 
\barr 
\frac{E_\Lambda \, d^3\sigma^{AB \to \Lambda X}}{d^3 \bfp_\Lambda} \>  
P_\Lambda &=& 
\sum_{a,b,c,d} \int \frac{dx_a \, dx_b \, dz}{\pi z^2} \> d^2\bfk_\perp \>  
f_{a/A}(x_a) \> f_{b/B}(x_b) \nonumber \\ 
&\times& \hat s \> \delta(\hat s + \hat t + \hat u) \>  
\frac{d\hat\sigma^{ab \to cd}}{d\hat t}(x_a, x_b, \bfk_\perp) \>  
\Delta^ND_{\Lup/c}(z, \bfk_\perp) \label{phgen} 
\earr 
where $\hat s, \> \hat t$ and $\hat u$ are the Mandelstam variables for the  
elementary process: for collinear configurations $\hat s = x_a x_b s, \>  
\hat t = x_at/z$ and $\hat u = x_bu/z$ and the modifications due to  
intrinsic $\bfk_\perp$ will be taken into account in the numerical  
evaluations. $E_\Lambda \, d^3\sigma^{AB \to \Lambda X}/d^3 \bfp_\Lambda$ is  
the unpolarized cross-section: 
\barr 
\frac{E_\Lambda \, d^3\sigma^{AB \to \Lambda X}}{d^3 \bfp_\Lambda} &=&   
\sum_{a,b,c,d} \int \frac{dx_a \, dx_b \, dz}{\pi z^2} \> d^2\bfk_\perp \>  
f_{a/A}(x_a) \> f_{b/B}(x_b) \nonumber \\ 
&\times& \hat s \> \delta(\hat s + \hat t + \hat u) \> 
\frac{d\hat\sigma^{ab \to cd}}{d\hat t}(x_a, x_b, \bfk_\perp) \>  
\hat D_{\Lambda/c}(z, k_\perp) \>. \label{unp} 
\earr 

In Eq.\ (\ref{phgen}) $\bfk_\perp$ is considered only where its absence 
would lead to zero polarization: that is, leading collinear configurations 
are assumed for partons $a$ and $b$ inside unpolarized hadrons $A$ and $B$, 
while transverse motion is considered in the fragmentation process. 
The final hadron, detected with momentum $\bfp_\Lambda$, is generated by  
the hadronization of a parton $c$ whose momentum, $\bfp_c = (\bfp_\Lambda 
- \bfk_\perp)/z$, varies with $\bfk_\perp$; also the corresponding 
elementary process, $ab \to cd$, depends on $\bfk_\perp$.  
 
$P_\Lambda$ is a function of the hyperon momentum $\bfp_\Lambda = \bfp_L +  
\bfp_T$ and is normally measured in the $AB$ c.m. frame as a function of  
$x_F \equiv 2p_L/\sqrt s$ and $p_T$. 
 
Notice that, in principle, there might be another contribution to  
the polarization of a final hadron produced at large $p_T$ in the high  
energy collision of two unpolarized hadrons; in analogy to   
Sivers' effect \cite{siv, noi1} one might introduce a new spin and 
$\bfk_\perp$ dependent distribution function:  
\barr 
\Delta^N f_{\aup/A}(x_a, \bfk_{\perp a}) &\equiv& 
\hat f_{\aup/A}(x_a, \bfk_{\perp a}) - \hat f_{\adown/A}(x_a, \bfk_{\perp a})  
\label{delf1}\\ 
&=& \hat f_{\aup/A}(x_a, \bfk_{\perp a}) 
- \hat f_{\aup/A}(x_a, - \bfk_{\perp a}) \>, 
\nonumber 
\earr 
{\it i.e.}\ the difference between the density numbers  
$\hat f_{\aup/A}(x_a, \bfk_{\perp a})$ and  
$\hat f_{\adown/A}(x_a,$ $\bfk_{\perp a})$ 
of partons $a$, with longitudinal momentum fraction $x_a$, transverse  
momentum $\bfk_{\perp a}$ and {\it transverse polarization} $\uparrow$ or  
$\downarrow$, inside an {\it unpolarized} hadron $A$.  
 
This idea was first applied to unpolarized lepto-production \cite{dan1} and  
to single spin asymmetries in $p \, p^\uparrow$ scattering \cite{dan3};   
the corresponding function, related to  
$\Delta^N f_{\aup/A}(x_a, \bfk_{\perp a})$, was denoted by $h_1^\perp$. 
In the present case of transversely polarized $\Lambda$ production this 
function would enter the cross-section accompanied by the transversity 
fragmentation function $H_1(z)$ (or $\Delta D_{h^\uparrow/a^\uparrow}$). 
We shall not consider this contribution here; not only because of the problems 
concerning $\Delta^N f_{\aup/A}(x_a, \bfk_{\perp a})$ discussed below,  
but also because the experimental evidence 
that the hyperon polarization is somewhat independent of the nature 
of the hadronic target suggests that the mechanism responsible for   
the polarization is in the hadronization process\footnote{This is not in 
contradiction with the observed spin transfer  
($D_{NN}$) in $p \, p^\uparrow \to \Lup \, X$ \cite{Bravar-97}, which in  
the factorized approach can be described in terms of the 
ordinary transversity distribution and fragmentation functions, rather than in 
terms of $\Delta^N D_{\hup/a}$.}. A clean test of  
this should come from a measurement of $P_\Lambda$ in unpolarized DIS  
processes, $\ell p \to \Lup X$ \cite{prep}.    
 
The $\bfk_\perp$ dependent functions considered in Refs.\ \cite{siv, noi1, 
noi3, dan1, col, Mulders-Tangerman-96} ($\Delta^N f_{a/A^\uparrow}$,  
$\Delta^N f_{\aup/A}$, $\Delta^N D_{h/a^\uparrow}$ and $\Delta^N D_{\hup/a}$, 
or, respectively,  
$f_{1T}^\perp$, $h_1^\perp$, $H_1^\perp$ and $D_{1T}^\perp$) have the 
common feature that the transverse momentum direction is correlated with the  
direction of the transverse spin of either a quark or a hadron, via a  
$\sin \phi$ dependence, as can be seen from Eq.\ (\ref{D1Tpdef}) for example.  
The reason for considering these functions is that this ``handedness'' of the  
transverse momentum compared to the 
transverse spin is displayed by the single spin asymmetry data  
in, for instance, $p \, p^\uparrow \to \pi \, X$ and $p \, p \to 
\Lambda^\uparrow \, X$.  
However, the problem of using such functions is that  
naively they appear to be absent due to time reversal invariance. This  
conclusion would be valid if the hadronic state appearing in the definition  
of such functions is treated as a plane wave state. One can then show that  
the functions are odd under the application of time reversal 
invariance, whereas hermiticity requires them to be even. 
If, however, initial or final state interactions are present, 
then time reversal symmetry will not prevent the appearance of nonzero  
(naively) T-odd functions. In the case of a state like  
$\vert P_h,S_h; X \rangle$ final state interactions are certainly present and  
nonzero fragmentation functions $\Delta^N D_{h/a^\uparrow}$ and  
$\Delta^N D_{\hup/a}$ are expected. However, for distribution functions  
this issue poses severe problems and since we will only consider fragmentation 
functions here, we refer to Refs.\ \cite{col, noi1, dan1}  
for more detailed discussions on this topic. 
 
The main difference between the function $\Delta^N D_{h/a^\uparrow}$ as 
originally proposed by Collins, and the function under present investigation  
$\Delta^N D_{\hup/a}$, is that the former is a so-called chiral-odd function, 
which means that it couples quarks with left- and right-handed chiralities,  
whereas the latter function is chiral-even. Since the pQCD interactions 
conserve chirality, chiral-odd functions must always be accompanied by a  
mass term or appear in pairs.  
Both options restrict the accessibility of such functions and 
make them harder to be determined separately. On the other hand, the  
chiral-even fragmentation 
function can simply occur accompanied by the unpolarized (chiral-even) 
distribution functions, which are the best determined 
quantities, allowing for a much cleaner extraction of the fragmentation 
function itself. Moreover, since chiral-even functions can appear in charged 
current mediated processes (as opposed to chiral-odd functions), more methods 
of extraction are available \cite{dan2}.  
 
As it was studied in Ref.\ \cite{Schaefer-T-99} the Collins function  
$H_1^\perp$ (or $\Delta^N D_{h/a^\uparrow}$) satisfies a sum rule arising  
from momentum conservation in the transverse directions. The same holds  
for the other $\bfk_{\perp}$-odd, T-odd fragmentation function  
$D_{1T}^\perp$ \cite{dan4}, 
\beq 
\sum_h \int dz \int d k_\perp^{2} \; \frac{k_\perp^2}{z M_h} \> 
D_{1T}^{\perp}(z,k_\perp^{}) = 0, \label{sumr1} 
\eeq 
or, in terms of the function $\Delta^ND_{\hup/a}$,  
\beq 
\sum_h C_h\,M_h \equiv \sum_h \int dz \int d^2 \bfk_{\perp} \;   
k_{\perp} \; \sin\phi\; \Delta^N D_{\hup/a} (z,\bfk_{\perp}) = 0 \>,  
\label{sumr2}   
\eeq 
which is equivalent to Eq.\ (\ref{sumr1}) via Eq.\ (\ref{rel}). 
 
Notice that the above sum rule can be written as 
\beq 
\sum_h \int dz \int d^2 \bfk_{\perp} \;   
\bfk_{\perp} \; \hat D_{\hup/a} (z,\bfk_{\perp}) = 0 \>, 
\label{sumr3}   
\eeq 
and the same holds independently for $\hat D_{\hdown/a} (z,\bfk_{\perp})$.   
Eq.\ (\ref{sumr3}) has a clear nontrivial physical meaning: for each 
polarization direction ($\uparrow$ or $\downarrow$) the total  
transverse momentum carried by spin 1/2 hadrons\footnote{Strictly speaking, 
the sum over $h$ is over all hadrons that carry transverse polarization, which 
might be true also for higher spin hadrons.} is zero. 
 
The sum over hadrons prevents a straightforward application of the sum rule  
to the case of $\Lambda$ production alone. It can not be used as a constraint  
on the parameterization of the function to be fitted to the data. However,  
we note that the integral $C_h\,M_h$ for each hadron type $h$ is the same  
function of the energy scale (implicit in all expressions), apart from as yet  
unknown normalization. In this sense it closely resembles the tensor charge. 
In other words, the running of the functions are the same for any type of 
hadron and there is no mixing with other functions. The ratios for different 
types of hadrons are constants, which allow for checks of consistency between 
sets of data obtained at different energies, without the need for evolution. 
These constants are universal, if indeed the T-odd fragmentation 
functions are universal. This universality is the main point of 
interest here: one wants to see if $\Lambda$ polarization from unpolarized 
quark fragmentation is independent of the initial state, as it is implicit 
when writing down the factorized cross-sections Eqs.\ (\ref{phgen}) and  
(\ref{unp}). At the present time, this can not be verified due to lack of 
data, but some predictions can be given \cite{prep} that will allow tests  
of this universality.    
 
We only consider the quark fragmenting into a $\Lambda$ and not into 
other hyperons, like the $\Sigma^0$. The latter actually produces a 
significant amount (20-30\%) of the $\Lambda$'s via the decay $\Sigma^0 
\to \Lambda \gamma$. The reason we do not introduce a separate $\Sigma^0$ 
fragmentation function at this stage is that the factorized approach by 
itself does not address such a separation (it is about a generic spin-1/2 
hadron of type $h$), unless one introduces some additional input, like a 
model based on $SU(3)$, or unless one applies it to separate sets of data 
for each hyperon (which are not available yet). Apart from that, for each 
quark flavour such a $\Sigma^0$ fragmentation function would evolve in the 
same way as the $\Lambda$ fragmentation function, which implies that their 
relative fraction stays constant. In this way we can view the $\Lambda$ 
fragmentation function as an effective fragmentation function that 
includes the contamination due to $\Sigma^0$'s. Indeed, the fragmentation 
functions we shall use in next Section have been obtained from fits 
to inclusive $\Lambda$ productions in $e^+e^-$ processes, independently  
of their origin.  
 
At a later stage one might make the distinction that the $\Sigma^0$  
fragmentation is a different fraction of the effective $\Lambda$  
fragmentation function for different quark flavours. One can insert all 
this information with hindsight and correct for it, but the present 
approach cannot be used to acquire this information unless the data would 
clearly distinguish the $\Lambda$'s coming from $\Sigma^0$'s. Our approach 
of using an effective $\Lambda$ fragmentation function would be more  
problematic if the $\Sigma^0$ would decay into other final states than  
$\Lambda \gamma$ (which branching ratio happens to be 100\%): then only a part 
of the total $\Sigma^0$ fragmentation function would be included into the  
effective $\Lambda$ fragmentation function and this would be energy dependent. 
 
In the case of longitudinally polarized $\Lambda$ production the 
$\Sigma^0\to \Lambda \gamma$ background gives rise to a depolarizing effect  
of about 10\% \cite{Burkardt-Jaffe}, but in the 
present situation of transversely polarized $\Lambda$ production this is 
not the case, since the photon does not carry away any of 
the transverse polarization and it hardly affects the definition of the plane  
compared to which the transverse polarization 
is measured. Therefore, the 
$\Sigma^0 \to \Lambda \gamma$ background does not produce a significant  
depolarizing effect for the transverse $\Lambda$ polarization and an effective 
$\Lambda$ polarizing fragmentation function can be used also.  
 
We shall now consider both $\Lambda$ and $\bar\Lambda$ production and 
attempt a determination of the polarizing fragmentation functions 
$\Delta^N D_{\Lambda^\uparrow/q}$ by comparing results for $P_\Lambda$ 
and $P_{\bar\Lambda}$ from Eqs.\ (\ref{phgen}) and (\ref{unp}) with  
data \cite{pl0}-\cite{pl4}.  
\vskip 18pt 
\nd 
{\bf 2. Numerical fits of data on {\mbox{\boldmath $P_\Lambda$}} from 
{\mbox{\boldmath $pN \to \Lambda X$}} processes} 
\vskip 6pt 
Eq.\ (\ref{phgen}), for proton-nucleon processes, can be rewritten as: 
\barr 
&& \frac{E_\Lambda \,  
d^3\sigma^{pN \to \Lambda X}}{d^3 \bfp_\Lambda} \> P_\Lambda =  
\sum_{a,b,c,d} \int_{(+k_\perp)} \hskip-18pt d^2\bfk_\perp \>  
\Biggl[ 
\int^1_{z_{\rm min}} \hskip-12pt dz \int^1_{x_{a\,\rm min}} \hskip-16pt dx_a  
\> \frac{1}{\pi z}\>\frac {\bar x_b^2 s}{(-t \Phi_t)} \label{phgen2} \\   
&&\times\,f_{a/p}(x_a) \> f_{b/N}(\bar x_b) \>  
\frac{d\hat\sigma^{ab \to cd}}{d\hat t}(x_a, \bar x_b, \bfk_\perp) - 
\{\bfk_\perp \to -\bfk_\perp\} \Biggr]  
\> \Delta^ND_{\Lup/c}(z, \bfk_\perp) \>, \nonumber 
\earr 
which deserves several comments and some explanation of notations. 
 
\begin{itemize} 
\item 
In deriving Eq.\ (\ref{phgen2}) from Eq.\ (\ref{phgen}) we have used  
the fact that $\Delta^ND_{\Lup/c}(z, \bfk_\perp)$, Eq.\ (\ref{deld1}), is  
an odd function of $\bfk_\perp$; the $(+k_\perp)$ integration region  
of $\bfk_\perp$ runs over one half-plane of its components. 
 
\item 
The $x_b$ integration has been performed by exploiting the  
$\delta(\hat s + \hat t + \hat u)$ function; the resulting value of  
$x_b$ is given by 
\beq 
\bar x_b = - \frac{x_a t \, \Phi_t}{x_a z s + u \, \Phi_u} \> , 
\label{barxb} 
\eeq 
where $\Phi_t$ and $\Phi_u$ are defined below. 
\item 
$\bfk_\perp$ could have any direction in the plane perpendicular to $\bfp_c$; 
however, due to parity conservation in the hadronization process 
-- Eq.\ (\ref{lamfn}) -- the only $\bfk_\perp$ component contributing to the 
polarizing fragmentation function is that perpendicular to $\bfP_\Lambda$, 
{\it i.e.}\ the component lying in the production plane, the $x-z$ plane in 
our conventions. To simplify the kinematics we shall then consider  
only the leading contributions of $\bfk_\perp$ vectors in the $x-z$ 
plane.    
 
\item 
$s, \> t$ and $u$ are the Mandelstam variables for the $pN \to \Lambda X$ 
process; in the simple planar configuration discussed above ($\bfp_c$ and 
$\bfk_\perp$ both lying in the $x-z$ production plane) they are related to  
the corresponding variables for the elementary processes by: 
\barr 
\hat s &=& 2\,p_a \cdot p_b = x_a x_b s  \nonumber \\   
\hat t &=& - 2\,p_a \cdot p_c = (x_a/z) \, t \,\Phi_t(\pm k_\perp)  
\label{mmhat} \\ 
\hat u &=& - 2\,p_b \cdot p_c = (x_b/z) \, u \,\Phi_u(\pm k_\perp) \nonumber  
\earr 
with  
\barr 
t \,\Phi_t(\pm k_\perp) = g(k_\perp) \left\{ t \mp 2 k_\perp  
\frac{\sqrt{stu}}{t+u} - [1 - g(k_\perp)]\>\frac{t-u}2 \right\} \label{phit}\\ 
u \,\Phi_u(\pm k_\perp) = g(k_\perp) \left\{ u \pm 2 k_\perp  
\frac{\sqrt{stu}}{t+u} + [1 - g(k_\perp)]\>\frac{t-u}2 \right\} \label{phiu} 
\earr   
where $g(k_\perp) = \sqrt{1 - (k_\perp/p_\Lambda)^2}$ and  
where $\pm k_\perp$ refers respectively to the configuration in which  
$\bfk_\perp$ points to the left or to the right of $\bfp_c$. 
At leading order in $k_\perp/p_\Lambda$ one has: 
\beq 
\Phi_t(k_\perp) = 1 - \frac{k_\perp}{p_\Lambda} \sqrt{\frac ut}  
\quad\quad\quad  
\Phi_u(k_\perp) = 1 + \frac{k_\perp}{p_\Lambda} \sqrt{\frac tu} \> \cdot 
\label{ptu}  
\eeq 
  
\item 
The lower integration limits in Eq.\ (\ref{phgen2}) are given by: 
\beq  
x_{a\,{\rm min}} = - \frac{u \,\Phi_u(\pm k_\perp)} 
{zs + t \,\Phi_t(\pm k_\perp)} \quad\quad 
z \geq - \frac{t \,\Phi_t(\pm k_\perp) + u\,\Phi_u(\pm k_\perp)}{s}  
\> \cdot \label{intlim} 
\eeq 
Notice that the integration limits are slightly different for $+k_\perp$ 
and $-k_\perp$; when replacing $\bfk_\perp$ with $-\bfk_\perp$ inside the  
square bracket of Eq.\ (\ref{phgen2}), one should not forget to change 
accordingly also the $z$ and $x_a$ integration limits, and the value of 
$\bar x_b$, Eq.\ (\ref{barxb}), although the results are only marginally  
affected by this.    
 
\item 
Eq.\ (\ref{phgen2}) can be schematically written as  
\barr 
&& d\sigma^{pN \to \Lambda X} \> P_\Lambda =  
 d\sigma^{pN \to \Lup X} - d\sigma^{pN \to \Ldown X}   
= \sum_{a,b,c,d} f_{a/p}(x_a) \otimes  f_{b/N}(x_b) \nonumber \\ 
&\otimes& \!\!\!\! [d\hat\sigma^{ab \to cd}(x_a, x_b, \bfk_\perp)  
- d\hat\sigma^{ab \to cd}(x_a, x_b, -\bfk_\perp)] 
\otimes \Delta^ND_{\Lup/c}(z, \bfk_\perp)  \label{phsch} 
\earr 
which shows clearly that $P_\Lambda$ is a higher twist  
effect, despite the fact that the polarizing fragmentation function 
$\Delta^ND_{\hup/a}$ is a leading twist function: this is due to the  
difference in the square brackets, $[d\hat\sigma(+\bfk_\perp) -  
d\hat\sigma(-\bfk_\perp)] \sim k_\perp/p_T$, similarly to what happens for 
the single spin asymmetries in $\pup p \to \pi X$ \cite{noi1,noi3}. 
 
\item 
In the computation of the unpolarized cross-section  
$E_\Lambda \, d^3\sigma^{pN \to \Lambda X}/d^3 \bfp_\Lambda$ intrinsic 
transverse motion is significant only in limited phase space regions:  
we have checked that the values obtained in most of the kinematical regions of  
available data do not vary whether or not taking into account $\bfk_\perp$.  
Notice that when taking into account $\bfk_\perp$ the expression for  
$E_\Lambda \, d^3\sigma^{pN \to \Lambda X}/d^3 \bfp_\Lambda$ is the same 
as Eq.\ (\ref{phgen2}) with the $-$ inside the square brackets replaced by  
a $+$ and $\Delta^ND_{\Lup/c}(z, \bfk_\perp)$ replaced by  
$\hat D_{\Lambda/c}(z, k_\perp)$.  
 
\item 
When computing the cross-sections for scattering off nuclei, 
$p A \to \Lambda^\uparrow X$, for which plenty of data are available, 
we have adopted the most simple incoherent sum, neglecting nuclear effects. 
That is, for the scattering off a nucleus with A nucleons and Z protons 
we use: 
\beq 
d\sigma^{pA \to \Lambda X} = Z \, d\sigma^{pp \to \Lambda X} 
+ (A-Z) \, d\sigma^{pn \to \Lambda X} \>. \label{phn} 
\eeq 
We have checked that different ways of taking into account nuclear  
effects leave results for $P_\Lambda$ -- a ratio of cross-sections --  
almost unchanged. The partonic distribution functions in a neutron are  
obtained from the usual proton distribution functions by applying isospin  
invariance.  
  
\item 
Eq.\ (\ref{phgen2}) holds for any spin 1/2 baryon; we shall use it also  
for $\bar\Lambda$'s, using charge conjugation invariance to obtain 
$\bar q \to \bar\Lambda$ fragmentation properties from $q \to \Lambda$ ones, 
which implies $D_{\bar\Lambda/\bar q} = D_{\Lambda/q}$
and $\Delta^ND_{\bar\Lambda^\uparrow/\bar q} = \Delta^ND_{\Lup/q}$. 
\end{itemize} 
 
We now use Eq.\ (\ref{phgen2}) in order to see whether or not it can reproduce 
the data and in order to obtain information on the new polarizing   
fragmentation functions. To do so we introduce a simple parameterization  
for these functions and fix the parameters by fitting the existing 
data on $P_\Lambda$ and $P_{\bar\Lambda}$ \cite{pl0}-\cite{pl4}. 
 
We assume that $\Delta^ND_{\Lup/c}(z, \bfk_\perp)$ is strongly peaked 
around an average value $\bfk^0_\perp$ lying in the production plane, 
so that we can expect: 
\beq 
\int_{(+k_\perp)} \hskip-18pt d^2\bfk_\perp \>  
\Delta^ND_{\Lup/c}(z, \bfk_\perp) \> F(\bfk_\perp) 
\simeq \Delta_0^ND_{\Lup/c}(z, k^0_\perp) \> F(\bfk^0_\perp) \>. 
\label{delta} 
\eeq 
Consistently, since in this case $F(\bfk_\perp)$ depends weakly on 
$\bfk_\perp$ when $k_\perp \ll p_T$, in the computation of the  
unpolarized cross-section we use:  
\beq 
\int_{(+k_\perp)} \hskip-18pt d^2\bfk_\perp \>  
\hat D_{\Lambda/c}(z, k_\perp) \> F(\bfk_\perp) 
\simeq  \frac{1}{2}\>D_{\Lambda/c}(z) \> F(\bfk^0_\perp) \>. 
\label{dzf} 
\eeq 
The average $k^0_\perp$ value depends on $z$ and we parameterize this  
dependence in a most natural way: 
\beq 
\frac {k^0_\perp(z)}{M} = K \> z^a (1-z)^b \>, \label{park} 
\eeq 
where $M$ is a momentum scale ($M = 1$ GeV/$c$); in performing the fit 
we demand that $K$, $a$ and $b$ are constrained so that they satisfy the  
kinematical bound $p_q^2 = (p_\Lambda^2 - k_\perp^2)/z^2 \ge p_\Lambda^2$,  
from which $k_\perp^2 \le (1-z^2)\,p_\Lambda^2$ and  
\beq 
k_\perp^0(z) \simordertwo (p_\Lambda)_{\rm min} \> \sqrt {1-z} \simeq  
(1 \, {\rm GeV}/c) \> \sqrt {1-z} \> , \label{bound} 
\eeq 
which implies $a \ge 0$ and $b \ge 0.5$. The values of $K$, $a$ and $b$ 
resulting from our best fit will have a clear physical meaning. 
 
We parameterize $\Delta_0^ND_{\Lup/c}(z, k^0_\perp)$ in a similar simple  
form: we know that it has to be zero when $k_\perp = 0$ and $z=1$; 
in addition, the positivity of the fragmentation  
functions -- Eq.\ (\ref{lamfn}) -- requires, at any  
$k_\perp$ value, $|\Delta^ND_{\hup/q}(z, k_\perp)| \leq  
\hat D_{h/q}(z, k_\perp)$. However, according to Eqs.~(\ref{delta}),  
(\ref{dzf}) and to take into account the mentioned [see comment  
after Eq.\ (\ref{intlim})] difference of the integration regions  
$(+k_\perp)$ and $(-k_\perp)$, which is significant    
at the boundaries of the kinematical ranges (when $p_T \simeq  
k_\perp$ and when $p_T \simeq p_{T\,\rm max}$) we prefer to impose the 
more stringent bound $|\Delta_0^ND_{\Lup/c}(z, k^0_\perp)| \leq 
D_{\Lambda/c}(z)/2$.   
Following Ref.\ \cite{bl}, this is automatically satisfied by taking: 
\barr 
\Delta_0^ND_{\Lup/q}(z, k^0_\perp) &=& N^{\,\prime}_q \>  
\frac{k^0_\perp(z)}{M} \> \left[ \frac{z^{\alpha_q}(1-z)^{\beta_q}} 
{c_q^{c_q} d_q^{d_q}/(c_q + d_q)^{c_q + d_q}} \right]  
\frac {D_{\Lambda/q}(z)}2 \nonumber \\ 
&=& N_q^{\,\prime} \, K \frac{z^{c_q}(1-z)^{d_q}} 
{c_q^{c_q} d_q^{d_q}/(c_q + d_q)^{c_q + d_q}} \>  
\frac {D_{\Lambda/q}(z)}2 \label{pard} \\ 
&\equiv& N_q \, z^{c_q}(1-z)^{d_q} \> \frac {D_{\Lambda/q}(z)}2 \>, 
\nonumber   
\earr 
where we have used Eq.\ (\ref{park}), and we require $c_q = a + \alpha_q > 0$,
$d_q = b + \beta_q > 0$, and $|N_q^{\,\prime}|\,K \leq 1$.  
 
We are now almost fully equipped to compute $P_\Lambda$ and 
$P_{\bar\Lambda}$; let us briefly 
discuss the remaining quantities appearing in Eq.\ (\ref{phgen2}).  
 
\begin{enumerate} 
\item[-] 
We sum over all possible elementary interactions, computed at lowest order  
in pQCD. The polarizing fragmentation functions -- parameterized as in 
Eq.\ (\ref{pard}) -- are supposed to be non vanishing only for $\Lambda$ 
valence quarks, $u$, $d$ and $s$; similarly for $\bar\Lambda$ valence 
antiquarks $\bar u$, $\bar d$ and $\bar s$. All contributions to the
unpolarized fragmentation functions -- from quarks, antiquarks and gluons --
are taken into account.   
 
\item[-] 
We adopt the unpolarized distribution functions of MRST \cite{mrst}. 
We have explicitly checked that a different choice makes no difference 
in our conclusions. We fix the QCD hard scale of distribution (and 
fragmentation) functions at 2 (GeV$/c)^2$, corresponding to an  
average value $p_T \simeq 1.5$ GeV$/c$. Since the range of $p_T$ values 
of the data is rather limited, no evolution effect would be visible anyway.  
  
\item[-] 
We use the set of unpolarized fragmentation functions of Ref.\ \cite{ind},  
which allows a separate determination of $D_{\Lambda/q}$ and   
$D_{\bar\Lambda/q}$, and which includes $\Lambda$'s both directly and 
indirectly produced; it also differentiates between the $s$ quark 
contribution and the $u$ and $d$ ones: the non strange fragmentation  
functions $D_{\Lambda/u} = D_{\Lambda/d}$ are suppressed by an 
$SU(3)$ symmetry breaking factor $\lambda=0.07$ as compared to  
$D_{\Lambda/s}$. In our parameterization of   
$\Delta_0^ND_{\Lup/q}(z, k^0_\perp)$, Eq.~(\ref{pard}), we use the same 
$D_{\Lambda/q}$ as given in Ref.\ \cite{ind}, keeping the same parameters  
$\alpha_q$ and $\beta_q$ ($c_q$ and $d_q$) for all quark flavours, but  
allowing for different values of $N_u = N_d$ and $N_s$. 
 
\end{enumerate} 
 
We can now use pQCD dynamics, together with the chosen distribution  
and fragmentation functions, and the parameterized expressions of the 
polarizing fragmentation functions, in Eq.\ (\ref{phgen2}) to compute  
$P_\Lambda$ and $P_{\bar \Lambda}$; by comparing with data we obtain  
the best fit values of the parameters  
$K$, $a$, $b$, $N_u=N_d$, $N_s$, $c_q$ and $d_q$ introduced in  
Eqs.\ (\ref{park}) and (\ref{pard}). Notice that we remain with 7 free  
parameters, $c_q$ and $d_q$ being the same for all flavours. 
 
Our best fit results ($\chi^2$/d.o.f. = 1.57), taking into account all data  
\cite{pl0}-\cite{pl4}, are shown in Figs.~1-5.  
They correspond to the best fit parameter values: 
\beq  
K = 0.69 \quad\quad\quad a = 0.36  \quad\quad\quad b = 0.53 \label{par1}  
\eeq 
\beq  
N_u = - 4.30 \quad\quad N_s = 1.13 \quad\quad 
c_q = 6.58 \quad\quad\quad d_q = 0.67 \>.\label{par2}  
\eeq 
 
Let us comment in some details on our results. 
 
In Fig.~1 and 2 
we present our best fits to $P_\Lambda$ as a function of  
$p_T$ for different $x_F$ values, as indicated in the figures: the famous 
approximately flat $p_T$ dependence, for $p_T$ greater than 1 GeV/$c$, is 
well reproduced. Such a behaviour, as expected, does not continue 
indefinitely with $p_T$ and we have explicitly checked that at larger values 
of $p_T$ the values of $P_\Lambda$ drop to zero: the shape of such a decrease,
contrary to what happens in the $p_T$ region of the data shown here, 
strongly depends on the assumptions about the nuclear corrections.     
It may be interesting to note that this fall-off has not yet been observed 
experimentally, but is expected to be first seen in the near-future  
BNL-RHIC data.  
 
Also the increase of $|P_\Lambda|$ with $x_F$ at fixed $p_T$ values 
can be well described, as shown in Fig.~3;  
the two curves correspond to  
$p_T = 1.5$ GeV/$c$ (solid) and $p_T = 3$ GeV/$c$ (dashed-dotted).  
 
The best fits of Figs.~1 and 2 are compared to $p$--$Be$  
data \cite{pl1}-\cite{pl4}; these are collected at two different energies,  
in the $p$--$Be$ c.m. frame, $\sqrt s \simeq 82$ GeV \cite{pl1}-\cite{pl3}  
and $\sqrt s \simeq 116$ GeV \cite{pl4}. Our calculations are performed at  
$\sqrt s = 80$ GeV; we have explicitly checked that by varying the  
energy between 80 and 120 GeV, our results for $P_\Lambda$ vary, in the  
kinematical range considered here, at most by 10\%, in agreement with the  
observed energy independence of the data. 
 
Some data from $p$--$p$ collisions are also available; in Ref.\ \cite{pl0} 
a linear fit to $P_\Lambda(x_F)$ is performed, collecting all data with  
$p_T \ge 0.96 $ GeV/$c$, for an average value $\langle p_T \rangle =  
1.1$ GeV/$c$. In Fig.~4 
we show these data and the linear fit (central line);  
the upper and lower lines show the fit error band. The solid  
line is our computation, at $p_T = 1.1$ GeV/$c$,  
with all parameters fixed as in Eqs.\ (\ref{par1}) and (\ref{par2}): our fit  
reproduces the data with good accuracy.  
 
In Fig.~5  
we show our best fit results for $P_{\bar\Lambda}$ as a function  
of $p_T$ for different $x_F$ values, as indicated in the figure: in this  
case all data \cite{pl1, pl3} are compatible with zero, with large errors,  
and the measured $x_F$ range is not as wide as for $P_\Lambda$, as 
expected from the lack of overlapping between $\bar\Lambda$ and nucleon 
valence quarks.
   
The resulting values of the parameters, Eqs.\ (\ref{par1}) and (\ref{par2}), 
are very realistic; notice in particular that $b$ essentially reaches its 
kinematical limit 0.5 and the whole function (\ref{park}) giving the average 
$k_\perp$ value of a $\Lambda$ inside a jet turns out to be very reasonable.  
 
Mostly $u$ and $d$ quarks 
contribute to $P_\Lambda$, resulting in a negative value of $N_u$; instead, 
$u$, $d$ and $s$ quarks all contribute significantly to $P_{\bar\Lambda}$ 
and opposite signs for $N_u$ and $N_s$ are found, inducing cancellations.  
 
In Fig.~6   
we plot $|\Delta_0^ND_{\Lup/u,d}|$ and $\Delta_0^ND_{\Lup/s}$ as given by 
Eq.\ (\ref{pard}) with the best fit parameters (\ref{par1}) and (\ref{par2}). 
We show, for comparison, also $D_{\Lambda/u,d}$ and $D_{\Lambda/s}$:  
notice how a tiny value of the polarizing fragmentation functions, in a  
limited large $z$ region, is enough to allow a good description of the data.  
This also shows that taking into account only valence quark contributions
to $\Delta^ND_{\Lup/q}$ -- as we have done -- is a justified assumption.
 
A different set of unpolarized fragmentation functions  
can be found in the literature \cite{defl}: it holds for the quark  
fragmentation into $\Lambda + \bar \Lambda$ and gives  
$D_{(\Lambda + \bar\Lambda)/q}$ rather than a separate expression  
of $D_{\Lambda/q}$ and $D_{\bar\Lambda/q}$; it would be appropriate 
to compute the $\Lambda + \bar\Lambda$ single spin asymmetry: 
\barr 
P_{\Lambda + \bar\Lambda} &=& \frac  
{d\sigma^{\Lup} + d\sigma^{\Blup} - d\sigma^{\Ldown} - d\sigma^{\Bldown}} 
{d\sigma^{\Lup} + d\sigma^{\Blup} + d\sigma^{\Ldown} + d\sigma^{\Bldown}}  
\label{plbl} \nonumber\\ 
&=& \left( P_\Lambda + \frac{d\sigma^{\bar\Lambda}}{d\sigma^\Lambda}  
P_{\bar\Lambda} \right) 
\left( 1 + \frac{d\sigma^{\bar\Lambda}}{d\sigma^\Lambda} \right)^{-1} \cdot 
\earr 
However, since one knows from experiments on $p-N$ reactions 
that in the kinematical range of interest  
$d\sigma^{\bar\Lambda} \ll d\sigma^\Lambda$ 
(and this is confirmed in our scheme, simply due to the dominance 
of $q$ over $\bar q$ in a nucleon), one can safely assume 
\beq 
P_{\Lambda + \bar\Lambda} \simeq P_\Lambda \>. \label{p=bp}  
\eeq   
This set -- differently from the one of Ref.\ \cite{ind} -- 
assumes $SU(3)$ symmetry and takes $D_{\Lambda/u} = D_{\Lambda/d} =  
D_{\Lambda/s}$.  
 
We have also computed $P_\Lambda$ with this second set of fragmentation  
functions; as in the previous case we have parameterized 
$\Delta^ND_{\Lup/q}$ according to Eq.\ (\ref{pard}), with the same values 
of $c_q$ and $d_q$ for each flavour, but different values 
of $N_{u,d}$ and $N_s$. 
We can equally well ($\chi^2$/d.o.f. = 1.85) 
fit the data on $P_\Lambda$, obtaining fits almost   
indistinguishable from those of Figs.\ 1-4, with the best fit values:   
\beq  
K = 0.66 \quad\quad\quad a = 0.37  \quad\quad\quad b = 0.50 \label{par3}  
\eeq 
\beq  
N_u = - 28.13 \quad\quad N_s = 57.53 \quad\quad 
c_q = 11.64 \quad\quad\quad d_q = 1.23 \label{par4}  
\eeq 
Notice that, also in this case of $SU(3)$ symmetric fragmentation functions  
$D_{\Lambda/q}$, and using only data on $P_\Lambda$, one reaches similar  
conclusions about the polarizing fragmentation functions 
$\Delta^ND_{\Lup/q}$: $N_{u,d} \not= N_s$ and not only is there a difference 
in magnitude, but once more one finds negative values for 
$\Delta^N_0D_{\Lup/u,d}$ and positive ones for $\Delta^N_0D_{\Lup/s}$. 
This seems to be a well established general trend. Plots analogous to those 
of Fig.~6, would show also in this case, Eqs. (\ref{pard}) and (\ref{par4}),
$\Delta^N_0D_{\Lup/s} > |\Delta^N_0D_{\Lup/u,d}|$. 
  
Very recently, new sets of quark and antiquark fragmentation functions 
into a $\Lambda$, based on a bag model calculation and a fit to 
$e^+e^-$ data, have been published \cite{blt}. Both a $SU(3)$ flavour 
symmetric and a $SU(3)$ symmetry breaking set are available, although
at a rather too low energy scale ($\mu = 0.25$ GeV). Nevertheless, 
we have also tried using these sets in our scheme to fit the data on 
$P_\Lambda$ and $P_{\bar\Lambda}$: once more we can fit the data, better 
with the symmetric than the asymmetric set, and, again, negative 
$\Delta^N_0D_{\Lup/u,d}$ and positive $\Delta^N_0D_{\Lup/s}$ are found,
with $\Delta^N_0D_{\Lup/s} > |\Delta^N_0D_{\Lup/u,d}|$.  
  
\vskip 18pt 
\nd 
{\bf 3. Conclusions} 
\vskip 6pt 
 
A phenomenological approach has been developed towards a consistent  
explanation and predictions of transverse single spin effects in processes  
with inclusively produced hadrons; we work in a kinematical region where  
pQCD and the factorization scheme can be used, but $p_T$ is not much larger  
than intrinsic $k_\perp$, so that higher twist contributions are still  
important. This applies to several processes for which data are available,  
like $\pup p \to \pi X$ \cite{art,noi3} and $pN \to \Lup X$. The single spin  
effect originates in the fragmentation process, either of a transversely  
polarized quark into an unpolarized hadron -- Collins' effect \cite{col} --  
or of an unpolarized quark into a transversely polarized hadron --  
the polarizing fragmentation functions \cite{Mulders-Tangerman-96}. 
Single spin effects in quark distribution functions \cite{siv} have also been 
discussed \cite{noi1,noi2}. 
      
We have considered here the well known and longstanding problem of the  
polarization of $\Lambda$ hyperons, produced at large $p_T$ in the  
collision of two unpolarized hadrons. We have assumed a generalized  
factorization scheme -- with the inclusion of intrinsic transverse  
motion -- with pQCD dynamics; the new, spin and $\bfk_\perp$ dependent, 
polarizing fragmentation functions $\Delta^ND_{\Lup/q}$ have been  
parameterized in a simple way and data on $p \, Be \to \Lup X$,  
$p \, Be \to \bar\Lambda^\uparrow X$ and $p\,p \to \Lup X$ have been used  
to determine the values of the parameters which give a best fit to the  
experimental measurements. 

The data can be described with remarkable accuracy in all their features:
the large negative values of the $\Lambda$ polarization, the increase of
its magnitude with $x_F$, the puzzling flat $p_T \simorder 1$ GeV/$c$ 
dependence and the $\sqrt s$ independence; data from $p$-$p$ processes are
in agreement with data from $p$-$Be$ interactions and also the 
tiny or zero values of $\bar\Lambda$ polarization are well reproduced. 
The resulting functions $\Delta^ND_{\Lup/q}$ are very reasonable and 
realistic.  
 
Different sets of unpolarized fragmentation functions -- either $SU(3)$ 
symmetric or not -- lead to very similar conclusions about these new 
polarizing fragmentation functions describing the hadronization  
process of an unpolarized quark into a polarized $\Lambda$: they have  
opposite signs for $u$ and $d$ quarks as compared with $s$ quarks and their 
magnitudes are larger for $s$ quarks. They are sizeable -- with respect to 
the unpolarized fragmentation functions -- only in limited $z$ regions, yet 
they can describe remarkably well the experiments.    
      
These polarizing fragmentation functions have a partonic interpretation and  
a formal definition, Eq.\ (\ref{D1Tpdef}); they are free from the ambiguities  
related to initial state interactions which might affect analogous  
distribution functions and we expect them to be universal, process independent 
functions. Our parameterization of $\Delta^ND_{\Lup/q}$ should allow us 
to give prediction for $\Lambda$ polarization in other processes; a study of  
$\ell p \to \Lup \, X$ and $e^+e^- \to \Lup \, X$ is in progress \cite{prep}.  
 
\vskip 18pt  
\nd   
{\bf Acknowledgements} 
\vskip 6pt
D.B. thanks RIKEN, Brookhaven National Laboratory and the U.S. Department of 
Energy for support under contract \# DE-AC02-98CH10886; M.A. thanks
RIKEN BNL Research Center for support during a Workshop last year, when 
this work was initiated.

\newpage 
 
\begin{figure}[c] 
 \begin{center} 
  \epsfig{file=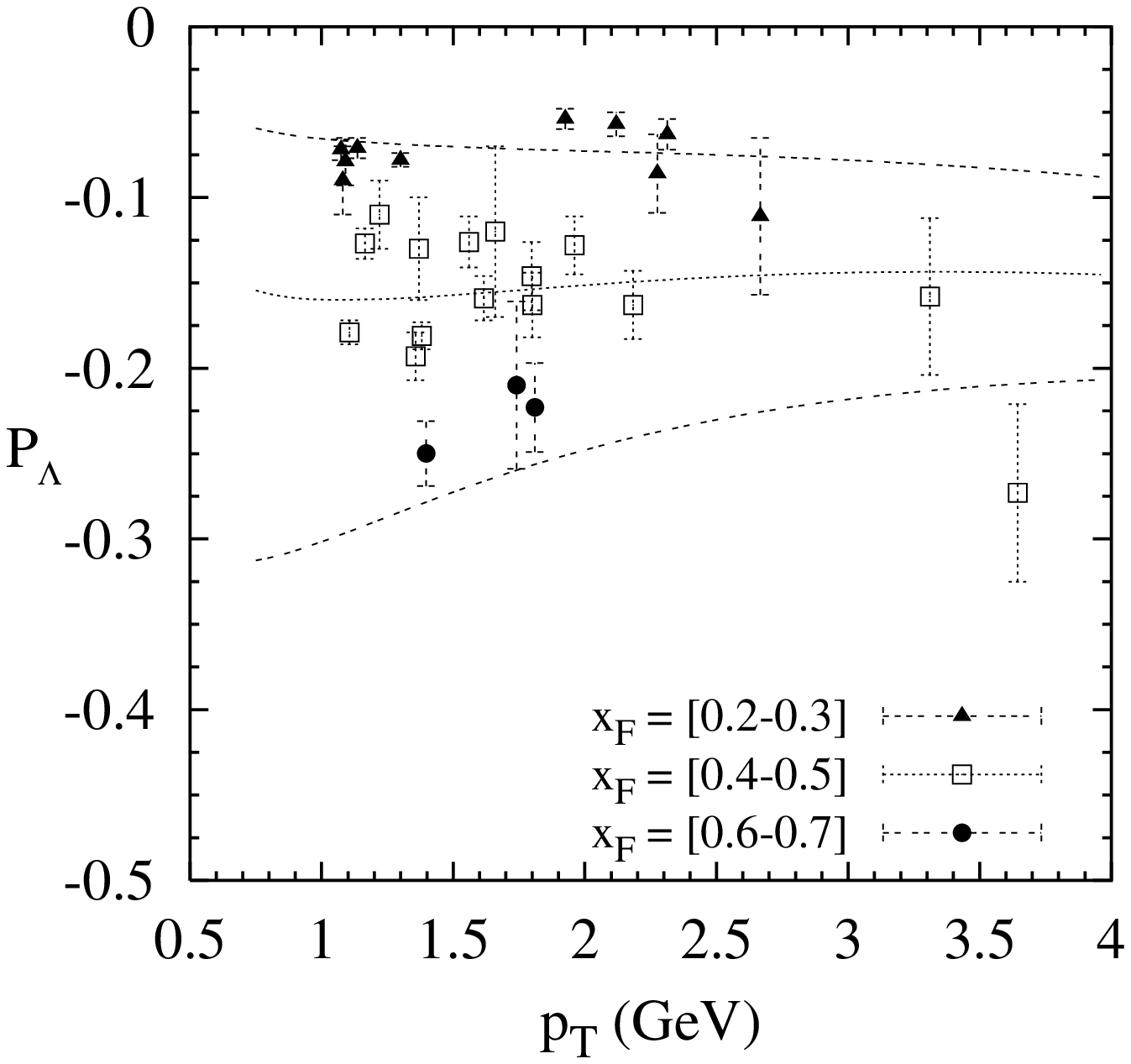,angle=-90,width=12.0cm} 
 \end{center} 
 \vspace{20pt} 
 \begin{center} 
  \begin{minipage}[c]{12cm} 
   {\small {\bf Fig.\ 1:} 
   Our best fit to $P_\Lambda$ data from $p$--$Be$ reactions, as a function  
   of $p_T$ and for different $x_F$ bins, as indicated in the figure. 
   Only some of the bins are shown; see Fig.~2 for complementary bins. 
   The experimental results, {\protect\cite{pl2}}-{\protect\cite{pl4}}, 
   are collected at two different c.m. energies, 
   $\sqrt{s}\simeq82$ GeV and $\sqrt{s}\simeq116$ GeV. 
   For each $x_F$-bin, the corresponding theoretical curve is evaluated 
   at the mean $x_F$ value in the bin, and at $\sqrt{s}=80$ GeV;  
   the results change very little with the energy. See the text for further  
   details. } 
  \end{minipage} 
 \end{center} 
\end{figure} 
 
\clearpage 
 
\begin{figure}[c] 
 \begin{center} 
  \epsfig{file=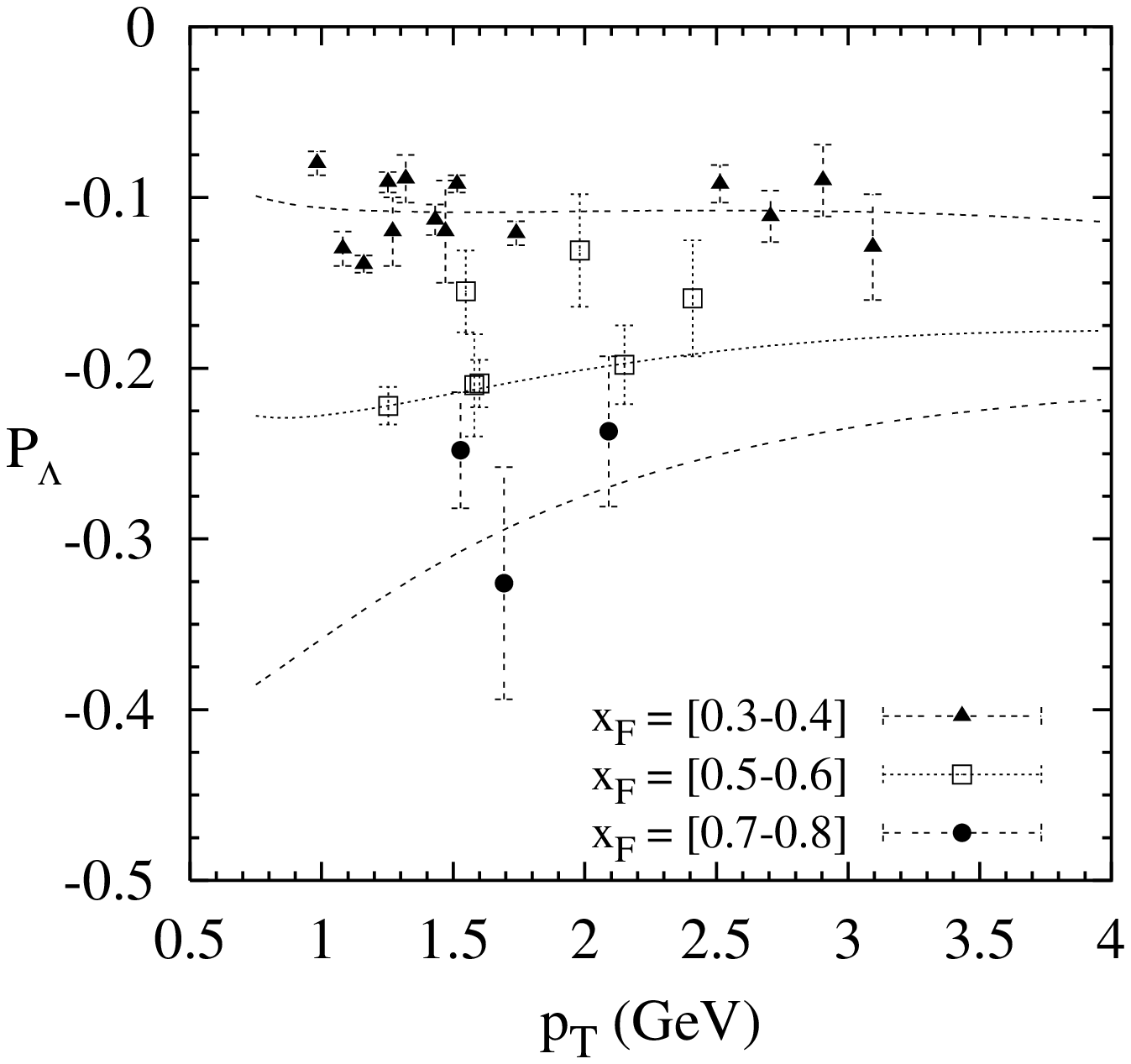,angle=-90,width=12.0cm} 
 \end{center} 
 \vspace{20pt} 
 \begin{center} 
  \begin{minipage}[c]{12cm} 
   {\small {\bf Fig.\ 2:} 
   Our best fit to $P_\Lambda$ data from $p$--$Be$ reactions, as a function  
   of $p_T$ and for different $x_F$ bins, as indicated in the figure. 
   Only some of the bins are shown; see Fig.~1 for complementary bins. 
   The experimental results,\protect\cite{pl2}-\protect\cite{pl4}, 
   are collected at two different c.m. energies, 
   $\sqrt{s}\simeq82$ GeV and $\sqrt{s}\simeq116$ GeV. 
   For each $x_F$-bin, the corresponding theoretical curve is evaluated 
   at the mean $x_F$ value in the bin, and at $\sqrt{s}=80$ GeV; 
   the results change very little with the energy. See the text for further  
   details. } 
  \end{minipage} 
 \end{center} 
\end{figure} 
 
\clearpage 
 
\begin{figure}[c] 
 \begin{center} 
  \epsfig{file=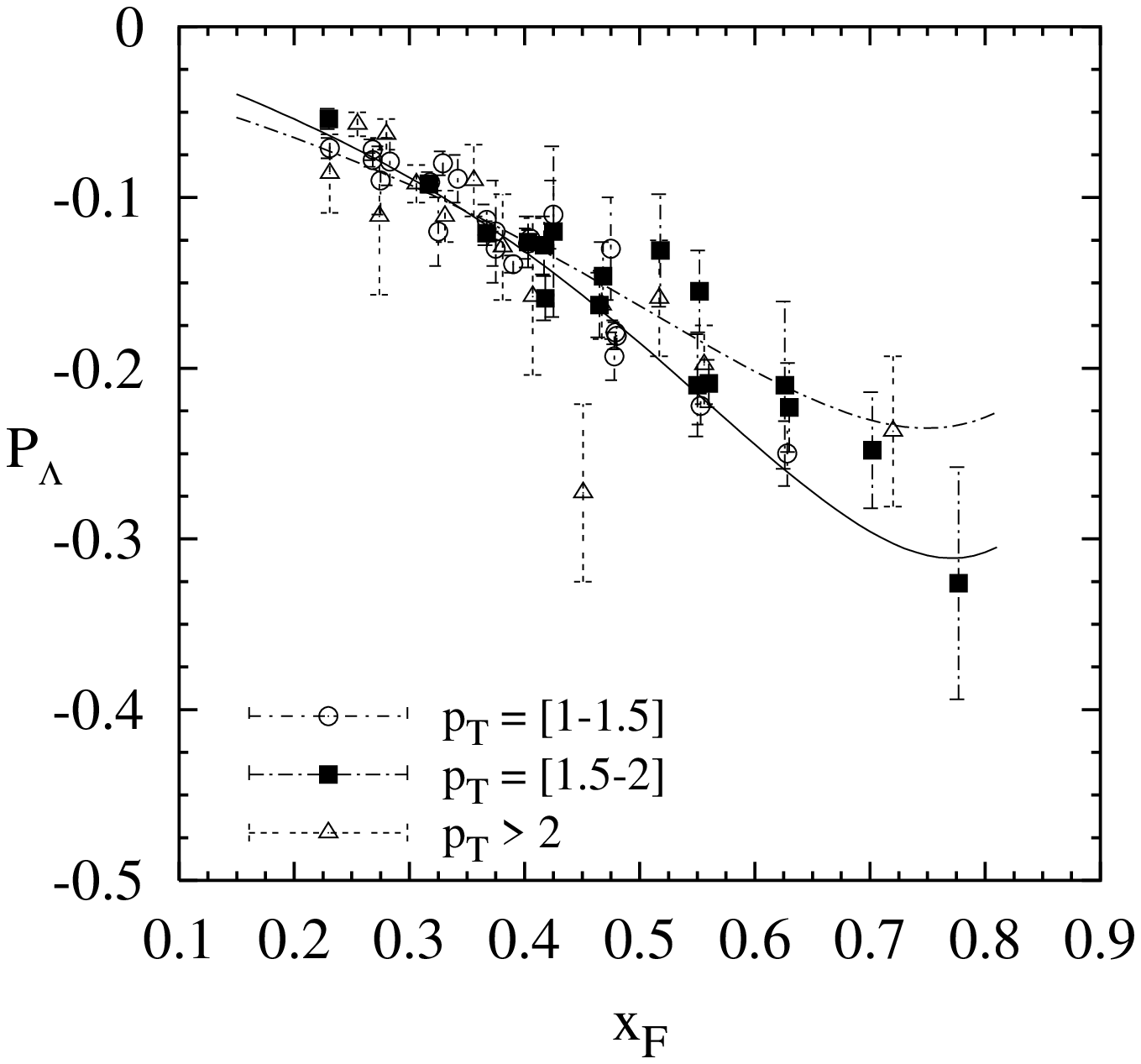,angle=-90,width=12.0cm} 
 \end{center} 
 \vspace{20pt} 
 \begin{center} 
  \begin{minipage}[c]{12cm} 
   {\small {\bf Fig.\ 3:} 
   $P_\Lambda$ data for $p$--$Be$ reactions, as a function of $x_F$ 
   and for different $p_T$ bins, as indicated in the figure. 
   The data are collected at two different c.m. energies, 
   $\sqrt{s}\simeq82$ GeV and $\sqrt{s}\simeq116$ GeV, 
   \protect\cite{pl2}-\protect\cite{pl4}. 
   The two theoretical curves, evaluated at $\sqrt{s}=80$ GeV, 
   correspond to $p_T=1.5$ GeV$/c$ (solid) and $p_T=3$ GeV$/c$ 
   (dot-dashed). } 
  \end{minipage} 
 \end{center} 
\end{figure} 

\clearpage 
 
\begin{figure}[c] 
 \begin{center} 
  \epsfig{file=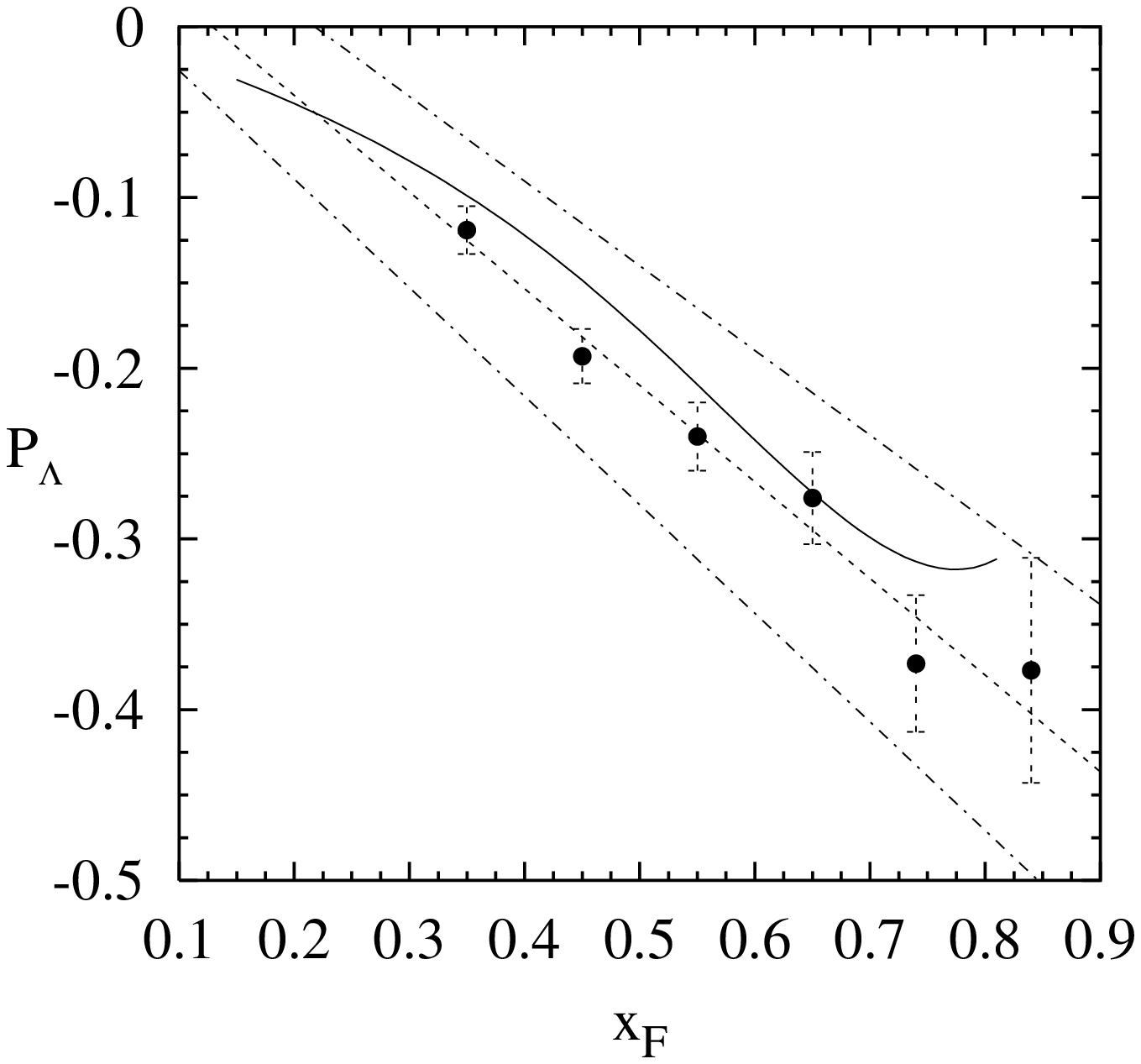,angle=-90,width=12.0cm} 
 \end{center} 
 \vspace{20pt} 
 \begin{center} 
  \begin{minipage}[c]{12cm} 
   {\small {\bf Fig.\ 4:} 
   Experimental results for $P_\Lambda$ in $p$--$p$ reactions, 
   as a function of $x_F$, from Ref.~\protect\cite{pl1}. 
   All data with $p_T \geq 0.96$ GeV$/c$ are collected, and 
   $\langle p_T\rangle=1.1$ GeV$/c$. Also shown is a linear fit 
   to the data, taken from Ref.~\protect\cite{pl1} (central line); 
   the upper and lower dot-dashed lines show the corresponding fit error band. 
   The solid curve shows the theoretical 
   computation, at $p_T = 1.1$ GeV/$c$, with all parameters fixed 
   as in Eqs.~(\protect\ref{par1}) and (\protect\ref{par2}). } 
  \end{minipage} 
 \end{center} 
\end{figure} 
 
\clearpage 
 
\begin{figure}[c] 
 \begin{center} 
  \epsfig{file=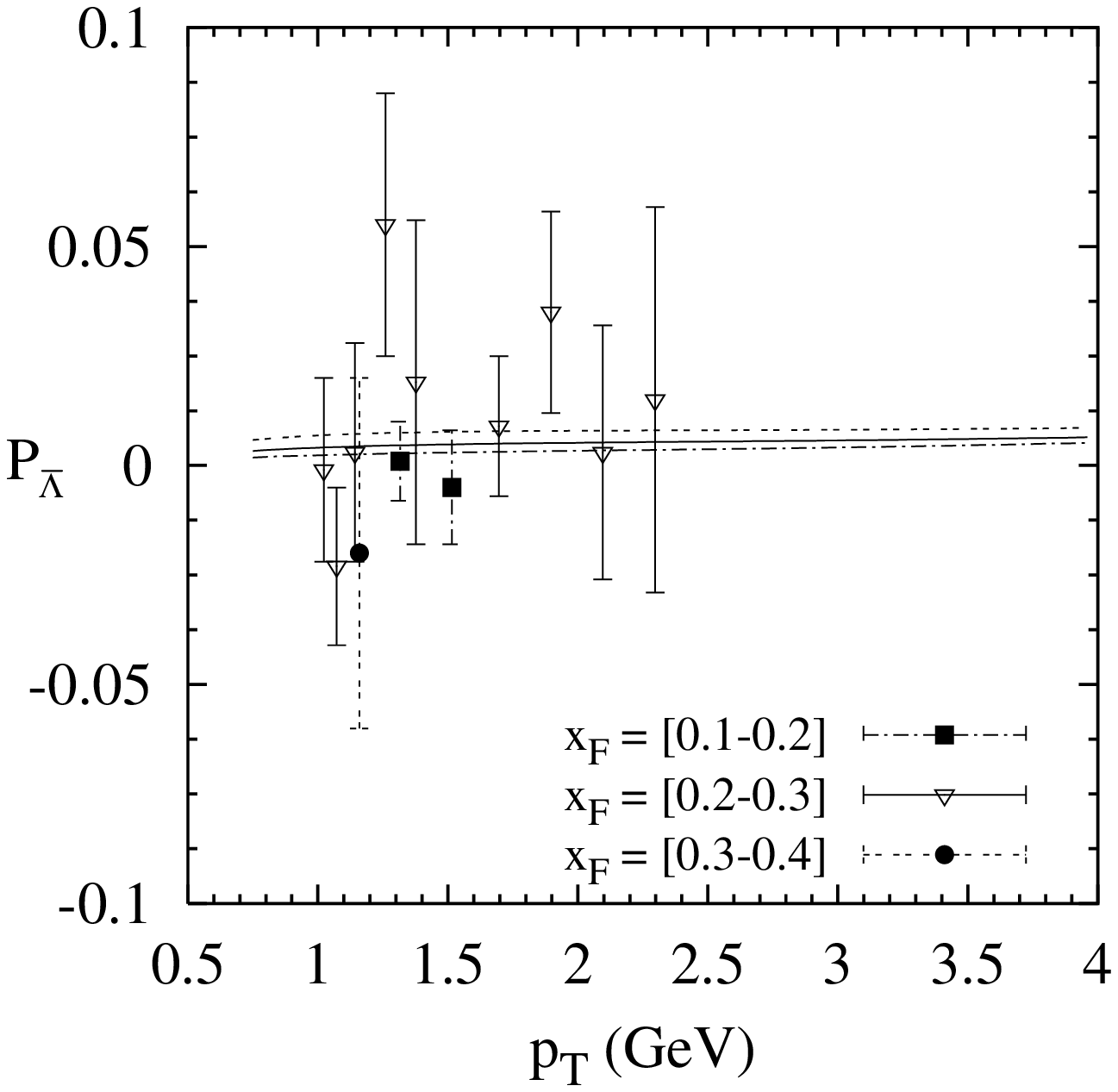,angle=-90,width=12.0cm} 
 \end{center} 
 \vspace{20pt} 
 \begin{center} 
  \begin{minipage}[c]{12cm} 
   {\small {\bf Fig.\ 5:} 
   Our best fit to $P_{\bar\Lambda}$ data from $p$--$Be$ reactions, as a 
   function of $p_T$ and for different $x_F$ bins, as indicated in the figure. 
   The experimental results \protect\cite{pl1, pl3} 
   are collected at the c.m. energies $\sqrt{s}\simeq82$ GeV. 
   For each $x_F$-bin, the corresponding theoretical curve is evaluated 
   at the mean $x_F$ value in the bin, and at $\sqrt{s}=80$ GeV; 
   the results change very little with the energy. } 
  \end{minipage} 
 \end{center} 
\end{figure} 
 
\clearpage 
 
\begin{figure}[c] 
 \begin{center} 
  \epsfig{file=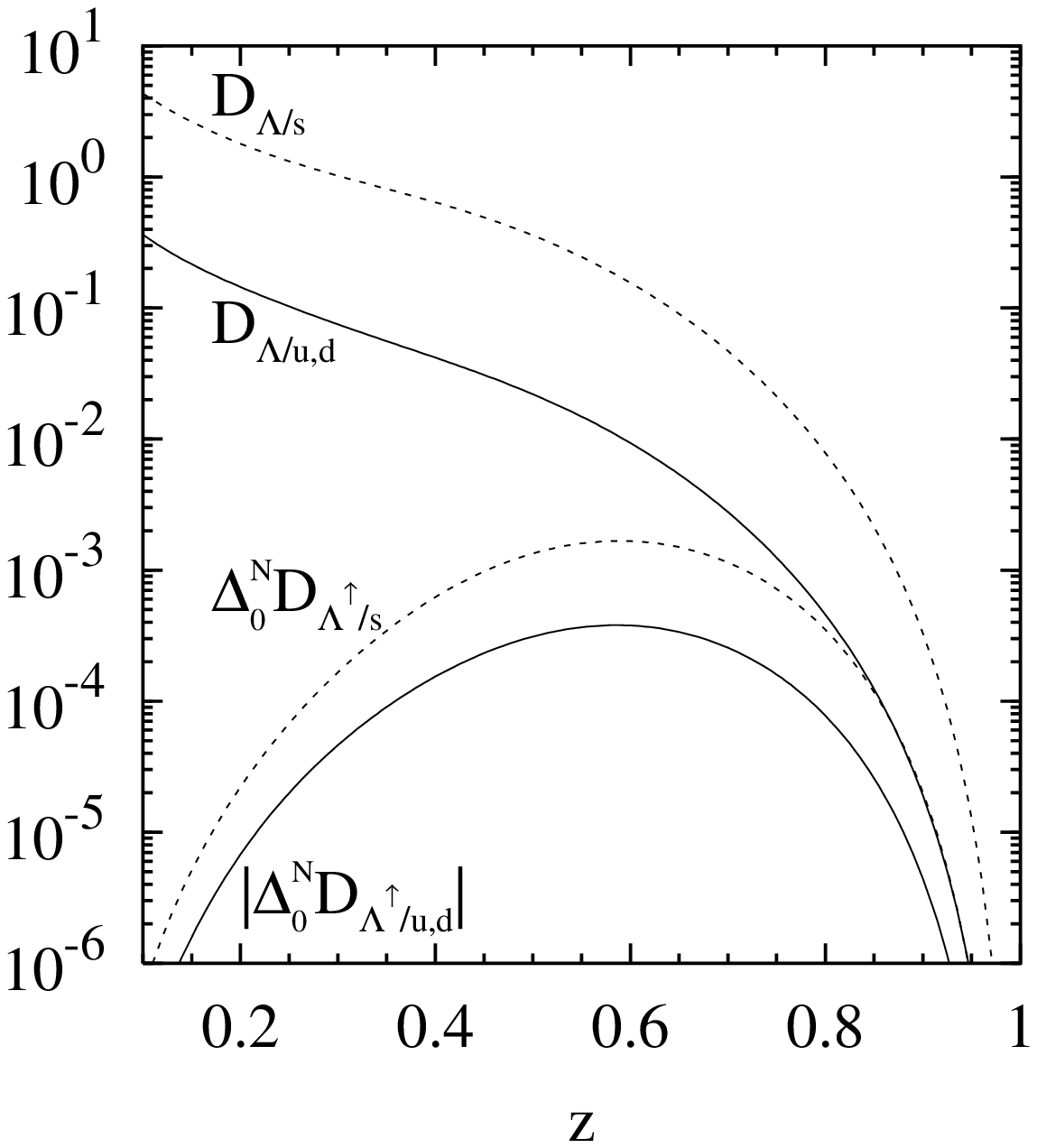,angle=-90,width=12.0cm} 
 \end{center} 
 \vspace{20pt} 
 \begin{center} 
  \begin{minipage}[c]{12cm} 
   {\small {\bf Fig.\ 6:} 
   Plot of $|\Delta_0^ND_{\Lup/u}|$ (= $|\Delta_0^ND_{\Lup/d}|$) and 
   $\Delta_0^ND_{\Lup/s}$, as given by Eq.~(\protect\ref{pard}) with 
   the best fit parameters 
   (\protect\ref{par1}) and (\protect\ref{par2}).
   For comparison we also show the unpolarized fragmentation functions
   $D_{\Lambda/u}$ (=$D_{\Lambda/d}$) and $D_{\Lambda/s}$ 
   \protect\cite{ind}. }
  \end{minipage} 
 \end{center} 
\end{figure} 
 
\end{document}